\documentclass[pdflatex,sn-mathphys-num]{sn-jnl}

\usepackage{graphicx}
\usepackage{dcolumn}
\usepackage{bm}
\usepackage{hyperref}
\usepackage{xr}
\usepackage{xr-hyper}
\usepackage{amsmath}

\usepackage[inline]{trackchanges}
\addeditor{YJ}

\makeatletter
\newcommand*{\addFileDependency}[1]{
  \typeout{(#1)}
  \@addtofilelist{#1}
  \IfFileExists{#1}{}{\typeout{No file #1.}}
}
\makeatother

\newcommand*{\myexternaldocument}[1]{%
    \externaldocument{#1}%
    \addFileDependency{#1.tex}%
    \addFileDependency{#1.aux}%
}

\myexternaldocument{./SI}

\usepackage{xcolor}
\usepackage{booktabs}
\usepackage{multirow}
\usepackage{siunitx}
\usepackage[mathlines]{lineno}

\theoremstyle{thmstyleone}%
\theoremstyle{thmstyletwo}%

\theoremstyle{thmstylethree}%

\begin{document}
\title{Expert-Grounded Automatic Prompt Engineering for Extracting Lattice Constants of High-Entropy Alloys from Scientific Publications using Large Language Models}



\author[1]{\fnm{Shunshun} \sur{Liu}}

\author[1]{\fnm{Talon R.} \sur{Booth}}

\author[2]{\fnm{Yangfeng} \sur{Ji}}
\author[3,4]{\fnm{Wesley} \sur{Reinhart}}

\author*[1]{\fnm{Prasanna V.} \sur{Balachandran}}\email{pvb5e@virginia.edu}

\affil*[1]{\orgdiv{Department of Materials Science and Engineering}, \orgname{University of Virginia}, \orgaddress{\city{Charlottesville}, \postcode{22903}, \state{VA}, \country{USA}}}

\affil[2]{\orgdiv{Department of Computer Science}, \orgname{University of Virginia}, \orgaddress{\city{Charlottesville}, \postcode{22903}, \state{VA}, \country{USA}}}

\affil[3]{\orgdiv{Department of Materials Science and Engineering}, \orgname{Pennsylvania State University}, \orgaddress{\city{State College}, \postcode{16802}, \state{PA}, \country{USA}}}

\affil[4]{\orgdiv{Institute for Computational and Data Sciences}, \orgname{Pennsylvania State University}, \orgaddress{\city{State College}, \postcode{16802}, \state{PA}, \country{USA}}}

\abstract{Large language models (LLMs) have shown promise for scientific data extraction from publications, but  rely on manual prompt refinement. We present an expert-grounded automatic prompt optimization framework that enhances LLM entity extraction reliability. Using high-entropy alloy lattice constant extraction as a testbed, we optimized prompts for Claude 3.5 Sonnet through feedback cycles on seven expert-annotated publications. Despite a modest optimization budget, recall improved from 0.27 to $>$~0.9, demonstrating that a small, expert-curated dataset can yield significant improvements. The approach was applied to extract lattice constants from 2,267 publications, yielding data for 1,861 compositions. The optimized prompt transferred effectively to newer models: Claude 4.5 Sonnet, GPT-5, and Gemini 2.5 Flash. Analysis revealed three categories of LLM mistakes: contextual hallucination, semantic misinterpretation, and unit conversion errors, emphasizing the need for validation protocols. These results establish feedback-guided prompt optimization as a low-cost, transferable methodology for reliable scientific data extraction, providing a scalable pathway for complex LLM-assisted research tasks.
}

\keywords{Large language models, prompt optimization, high-entropy alloys, lattice constants, scientific data extraction}



\maketitle

\section*{Introduction}\label{sec:intro}
Materials informatics studies based on experimental data have historically relied on manual data extraction from scientific publications. This process is tedious and time-consuming. Consequently, scaling this approach to extract data from thousands of publications remains impractical. Over the past decade, researchers have focused on automating data extraction from unstructured sources, such as journal publications. Initially, natural language processing (NLP) techniques were developed that relied on expert annotation~\cite{kim2017materials, venugopal2019picture, olivetti2020data, shetty2021automated, wang2022automated}, and more recently, there has been a shift toward leveraging large language models (LLMs) for this purpose~\cite{zheng2023chatgpt, gilligan2023rule, sayeed2024annotating, dagdelen2024structured, Polak2024, jiang2025applications, foppiano2024mining, liu2024prompt, ekuma2025dynamic,  biswajeet2025leveraging, liu2025harnessing, turan2025revolutionizing, polak2025leveraging}. Our survey of LLM applications in materials science literature (Figure S1 in the Supplementary Information, SI) identified automated knowledge and data extraction as the most prevalent use cases. These studies have demonstrated that LLM-based data extraction from scientific publications can achieve reasonable accuracy, but performance depends critically on prompt quality. A prompt is a structured set of instructions provided to an LLM that specifies the task, defines constraints, and outlines output format and content \cite{white2023prompt, schulhoff2024prompt}. Effective prompts typically include four elements \cite{zheng2023chatgpt, Polak2024, white2023prompt, schulhoff2024prompt, giray2023prompt}: (1) \emph{Instructions} that define the task and desired behavior, (2) \emph{Context} that provides relevant background information, (3) \emph{Input Data} to be processed, and (4) \emph{Output Indicators} that specify the desired format and structure of the response. 

Our literature survey identified several recent publications that provide systematic approaches to prompt engineering for extracting materials science data from scientific publications. We discuss two representative studies in detail below. Zheng et al.~\cite{zheng2023chatgpt} developed a human-in-the-loop prompt engineering strategy for chemistry-focused applications based on three core principles: ($i$) well-formulated prompts designed to minimize hallucination by avoiding queries that may produce fabricated content, ($ii$) detailed instructions that convey context and the desired response format, and ($iii$) structured output requirements using well-defined templates to facilitate extraction. Their workflow first identified relevant sections in publications (those containing metal-organic framework, MOF, synthesis conditions), confirmed data presence, and then extracted MOF synthesis parameters from those sections. The utilization of GPT-3.5 and GPT-4, coupled with an interactive prompt refinement strategy enhanced by human feedback, demonstrated significant efficacy in achieving robust performance metrics, which the authors discussed in terms of the precision, recall, and F1 score. Polak and Morgan~\cite{Polak2024} developed a two-stage zero-shot prompt engineering strategy. Stage 1 employed a relevancy prompt to identify sentences containing target data, while Stage 2 used question-and-answer prompts to extract specific information. They demonstrated this approach by constructing two databases: critical cooling rates of metallic glasses and yield strengths of high-entropy alloys (HEAs). Using GPT-3.5 (gpt-3.5-turbo-0301), GPT-4 (gpt-4-0314), and LLaMA2-chat (70B), they also evaluated extraction accuracy through precision and recall metrics. Additional studies by Ha et al.~\cite{ha2025ai} (nanotoxicity), Zhang et al.~\cite{zhang2024gptarticleextractor} (magnetic materials), Liu et al.~\cite{liu2024prompt} (metallic glasses), Gupta et al.~\cite{gupta2024data} (polymers), and Prasad et al.~\cite{prasad2024towards} (titanium-based alloys) further demonstrated the value of carefully structured prompts for large-scale data extraction from scientific texts.

\begin{figure}[!t]
    \centering
    \includegraphics[width=\linewidth]{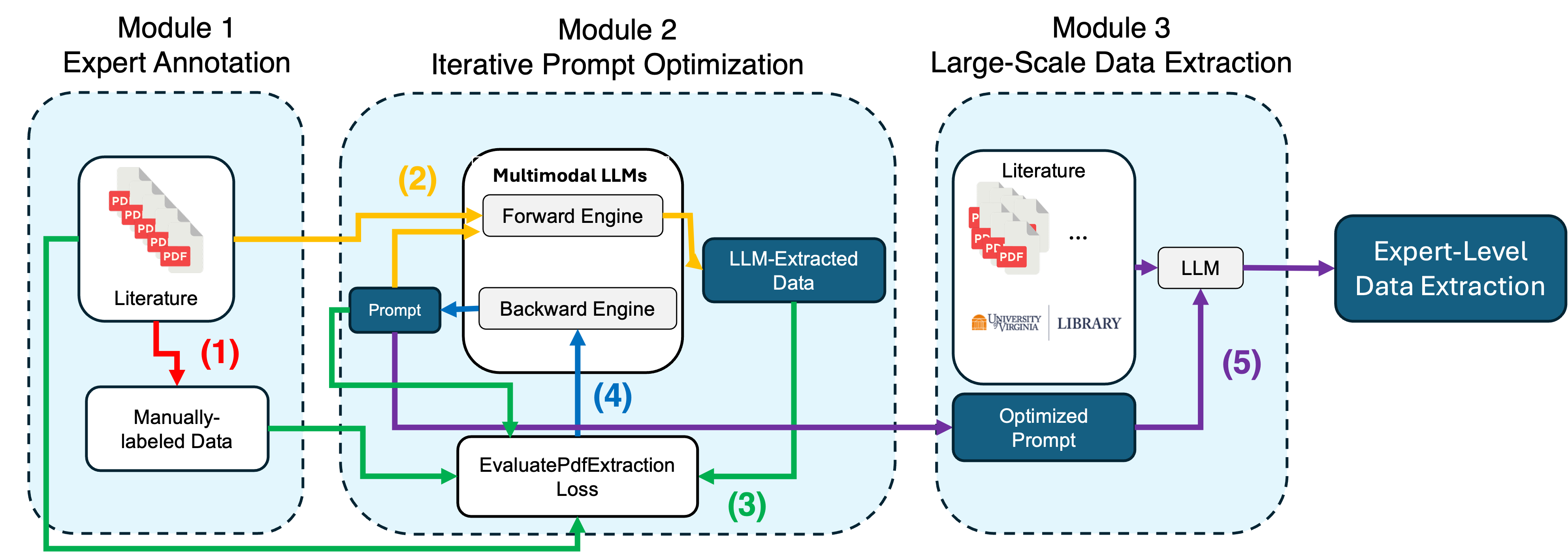}
    \caption{Schematic workflow for expert-grounded prompt optimization and large-scale data extraction. Numbers in parentheses indicate the sequential workflow steps, with corresponding arrow colors. The details of each module are described in the main text.}
    \label{fig:workflow}
\end{figure}

Current prompt engineering methods, despite advancements, have a significant limitation. They require considerable human involvement for refining prompts and are typically optimized based on individual publications rather than across a collection of publications. To address this limitation, we introduce a prompt engineering strategy that integrates domain expert-constructed data into an automated optimization process, rather than relying on post-hoc, human-in-the-loop refinement. Our workflow is shown in \autoref{fig:workflow}. Building on the \textsc{TextGrad} framework~\cite{yuksekgonul2025optimizing}, which enables prompt optimization through backpropagation of LLM-generated ``textual gradients'', our workflow leverages a small, expert-grounded dataset carefully curated from multiple scientific publications to automatically and iteratively refine prompts without human-in-the-loop.
These optimized prompts can then be used for large-scale automated data extraction from thousands of scientific publications. We demonstrate this methodology using HEAs as a test case, which are an emerging class of materials with significant technological interest~\cite{miracle2017critical, george2020high, han2024multifunctional}. In our prompts, we use the term ``high entropy alloys'', which the LLM interprets as encompassing multi-principal element alloys (MPEAs) and compositionally complex alloys (CCAs), as these terms are used interchangeably in the literature. For consistency, we use HEAs throughout this manuscript, with the understanding that it broadly encompasses MPEAs and/or CCAs. 

Our objective is to develop an expert-grounded, automatic prompt optimization framework for extracting experimentally measured, room temperature lattice constant data from thousands of HEA publications using LLMs. Current HEA databases, such as the one by Borg et al.~\cite{borg2020expanded}, which includes mechanical properties and phase information, and the database constructed by Polak and Morgan~\cite{Polak2024}, which contains yield strength data along with annotated source passages, do not provide lattice constant data for experimentally synthesized alloys. It is essential to construct an HEA dataset that includes lattice constant data from the experimental HEA literature to advance alloy development efforts ~\cite{tandoc2025bond}. They are important inputs for predicting solid-solution strengthening effects using mechanistic models~\cite{maresca2020mechanistic, liu2022integrating} and help estimate the coherency strain relationships between the matrix and precipitate phases in precipitation-hardened systems~\cite{li2023evaluating}. However, the manual extraction of lattice constant data from the rapidly expanding HEA scientific publications is impractical, creating a bottleneck for materials design and discovery efforts.

Our expert-grounded prompt optimization framework consists of three modules, as illustrated schematically in \autoref{fig:workflow}. Additional details can be found in the Methods section. Module 1 (Expert Annotation): Domain experts manually create a prompt training dataset by surveying a small subset of scientific publications. This step establishes the named entities to be extracted, the required data fields, and the structured format that guides subsequent prompt optimization. Module 2 (Iterative Prompt Optimization): The focus is on iterative prompt optimization using an LLM and \textsc{TextGrad} framework \cite{yuksekgonul2025optimizing}. This module uses the same publications that were annotated by domain experts in Module 1. It takes two inputs: ($i$) scientific publications in the PDF format that were used to construct the expert-annotated dataset from Module 1 and  ($ii$) an initial prompt. Through a multimodal LLM system and a workflow that utilizes forward and backward engines, the initial prompt is iteratively and automatically refined.
Claude 3.5 Sonnet \cite{anthropic2024claude} serves as both the forward engine for data extraction and the backward engine for generating improvement suggestions. The \texttt{EvaluatePdfExtractionLoss} function, implemented using the \textsc{TextGrad} loss module, compares LLM-extracted data against expert-labeled ground truth to compute ``textual gradients''. This optimization loop continues until the iteratively refined prompts yield satisfactory extraction performance. Module 3 (Large-Scale Data Extraction): The optimized prompts from Module 2 are deployed to extract data from thousands of scientific publications. This sequential, three module approach enables systematic prompt refinement using limited expert-curated data, followed by large-scale extraction across thousands of scientific literature. While this paper demonstrated the approach using Claude 3.5 Sonnet, any combination of commercial or open-source LLMs can also function as forward or backward engines.

Applying this approach to 2,267 scientific publications, we extracted lattice constant data for 1,861 HEA compositions. Beyond lattice constants, we also captured chemical compositions (both nominal and measured, if available), phase information, and processing techniques. Following established techniques in the literature, we evaluated extraction performance using precision, recall, and F1 score metrics. Analysis of these 1,861 entries identified 186 materials that exhibited single-phase body-centered cubic (BCC) structure in the as-cast processing condition. To demonstrate the utility of the extracted dataset for downstream applications, we built data-driven ML models that predict lattice constants as a function of as-cast, single-phase BCC HEA compositions, which are promising candidates for high-temperature structural applications~\cite{feng2021superior, lee2022explainable, karumuri2025design}. 

Although the Claude 3.5 Sonnet LLM excelled at large-scale data extraction, we also discuss three challenges that posed risks to extraction reliability: ($i$) contextual hallucination, where the model generated physically plausible but textually unsupported values \cite{zhang2025siren, huang2025survey}; ($ii$) semantic misinterpretation \cite{zhang2025siren, huang2025survey}, where technical terminology or implicit context was misunderstood; and ($iii$) unit inconsistencies that complicated downstream processing. An in-depth analysis of these limitations offers insight into the current challenges related to LLM-based data extraction from scientific literature.

Finally, while newer LLMs demonstrate strong zero-shot reasoning, certain scientific tasks, such as synthesizing information across modalities, remain challenging~\cite{polak2025leveraging, turan2025revolutionizing}. Consequently, systematic and automated prompt optimization procedures are essential for pushing model performance to its practical ceiling. Here, we have implemented such a procedure and validated it using HEA lattice constant extraction as a representative benchmark. While the prompt optimization workflow was based on the Claude 3.5 Sonnet model, the optimized prompt showed excellent transferability when tested on current LLM models, including Claude 4.5 Sonnet, Gemini 2.5 Flash, and GPT-5. Our results suggest a path forward for domain scientists in light of the rapidly advancing capability of frontier models.

\section*{Results}\label{sec:results}

\subsection*{Initial Prompt vs Optimized Prompt}
\begin{figure*}[ht]
  \centering
  \includegraphics[width=\linewidth]{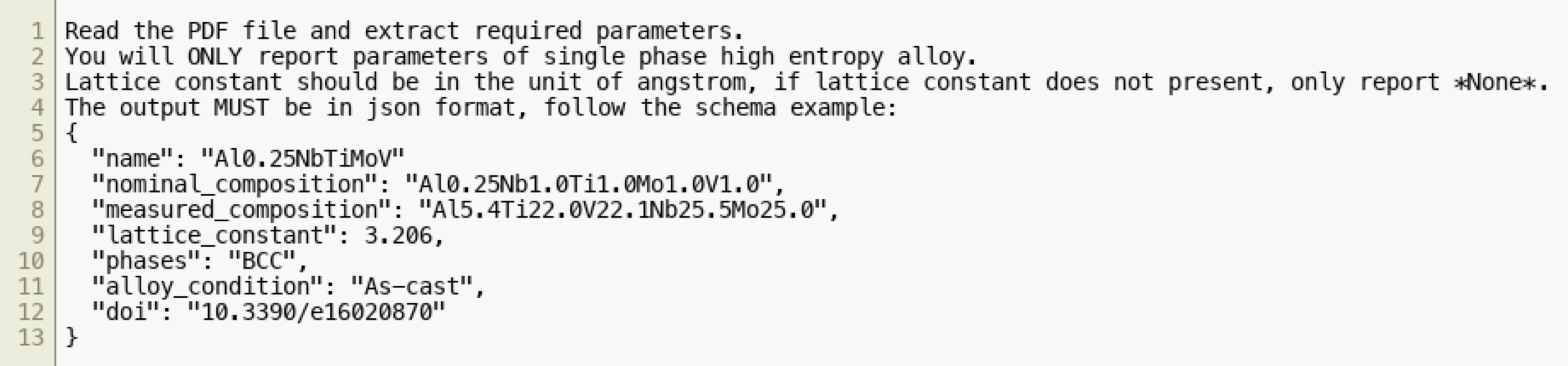}
  \caption{Full initial prompt. A machine-readable version is provided in the Figure S2 in SI.}
  \label{fig:initial_prompt}
\end{figure*}

\begin{figure*}[ht!]
  \centering
  \includegraphics[width=\linewidth]{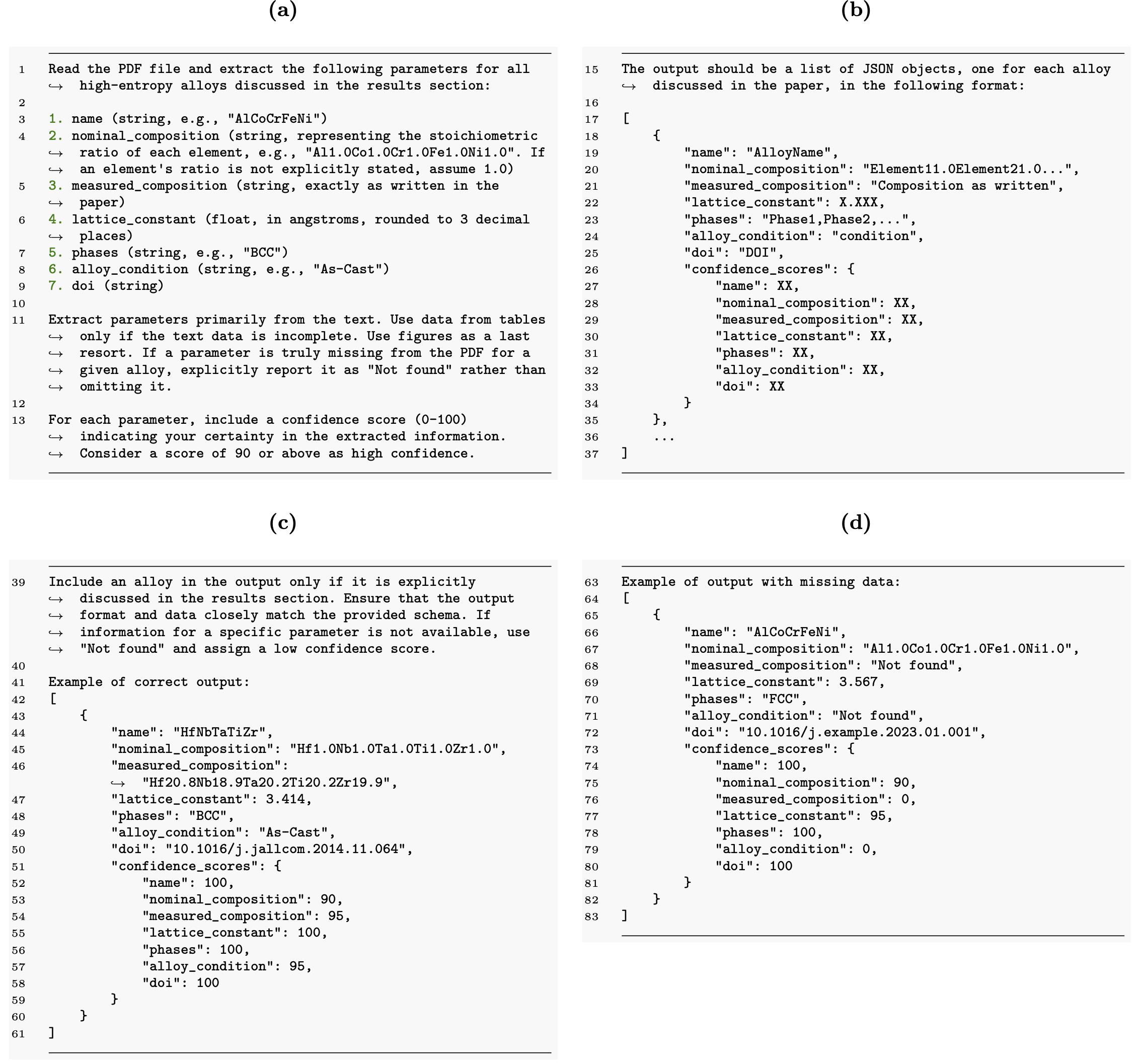}
  \caption{A condensed and streamlined visualization of the complete optimized prompt, broken down into four panels: (a) Task instructions, the scope of extraction, and the definition of the data extraction scheme. (b) An example of the extracted data in JSON format. (c) Additional instructions along with a JSON example demonstrating the correct output. (d) A JSON example illustrating how to handle missing values. A machine-readable version is provided in the Figure S3 in SI.}
  \label{fig:optimized_prompt}
\end{figure*}

The initial and optimized prompts are shown in \autoref{fig:initial_prompt} and \autoref{fig:optimized_prompt}, respectively. The initial prompt was constructed based on the domain expertise of the authors; no LLM consultation or systematic prompt engineering principles were employed in its design. Analysis of initial extraction attempts revealed three deficiencies that degraded output quality: ($i$) ambiguous instructions and incomplete extraction rules, ($ii$) inconsistent output formatting, and ($iii$) poorly defined extraction scope. We discuss each limitation and the corresponding improvements incorporated into the optimized prompt below.

One of the sources for ambiguity in the initial prompt stemmed from undefined technical terminology and missing handling rules. For example, the phrase ``single phase'' was not explicitly defined and there were no clear instructions provided for materials with multiple reported phases. The optimized prompt addresses this problem by broadening the scope to all HEAs (\autoref{fig:optimized_prompt}, line 1) while removing the ambiguous ``single phase'' constraint. This refinement is acceptable because single phase HEAs are a subset of the complex HEA phase space, which can be filtered during post-processing based on the extracted phase information. Additionally, the initial prompt lacked rules for handling missing data or multiple entries per publication.
The optimized prompt explicitly addresses these cases through both textual instructions (lines 11, 39) and concrete formatting examples: lines 15–37 demonstrate handling multiple material entries, while lines 63–83 illustrate proper missing value notation.

The second weakness involved output formatting. The initial prompt defined the data schema, but omitted formatting specifications for individual values. Without explicit guidance, the LLM was found to be prone to produce inconsistent outputs. The optimized prompt eliminates this weakness through detailed formatting requirements (lines 3–7), ensuring uniform outputs that match the desired schema.

The third weakness concerns undefined extraction scope. Without explicit boundaries, the LLM can extract spurious data from reference lists or citations, capturing materials mentioned but not experimentally investigated in the paper. To address this weakness, the optimized prompt establishes clear scope priorities (lines 11, 39): \textsf{``Extract information from results, focus on text first, then table, use figure at last.''}

\subsection*{Data Retrieval Performance on the Expert-Labeled Training Set}
We evaluated the data extraction performance of the initial and the optimized prompts on the expert-labeled training set, which included seven research papers containing data for 22 single-phase HEAs. The Claude 3.5 Sonnet model with the initial prompt was only able to extract seven entries from the seven publications. Extraction metrics for the initial prompt are given in \autoref{tab:compare_pr}, which reveal severe recall limitations: only 0.273 (27.3\%) of expert-identified HEA entries were successfully extracted. This indicates that approximately 73\% of relevant data remained undetected, highlighting the need for systematic prompt engineering. One of the reasons for the low recall is attributed to the incomplete extraction rules in the initial prompt. Specifically, the prompt implicitly restricted extraction to one HEA composition per publication, despite many papers reporting multiple HEA compositions. Among the seven extracted entries, one of the HEA compositions includes AlHfNbTaTiZr, which was taken from the work of Lin et al.~\cite{LIN2015100}. This HEA composition was deliberately excluded from our expert-constructed training set because the authors reported two different BCC lattice constants, suggesting a multi-phase microstructure instead of a single-phase BCC. We interpreted this as evidence of a two-phase BCC composition rather than the single-phase condition. With one incorrect extraction entry out of seven total extractions, the initial prompt achieved an overall precision of 0.857.

In contrast, the optimized prompt extracted 23 single-phase HEA composition entries from the seven publications, exceeding the 22 expert-labeled single-phase HEA compositions. Two out of 23 extracted entries were HEAs referenced in the papers, but not experimentally characterized in those studies, representing false positives (FPs). Additionally, the optimized prompt also extracted AlHfNbTaTiZr from Lin et al.~\cite{LIN2015100} and classified it as single-phase BCC, despite the two different lattice constants discussed earlier. 

The optimized prompt achieved a precision of 0.870 (20/23) and a recall of 0.909 (20/22) when evaluated against the specific single-phase HEA criterion. However, the instruction from the optimized prompt to extract \textsf{``all high-entropy alloys discussed in the results section''} expanded the effective scope beyond single-phase compositions. When evaluated against all HEA entries (regardless of single-phase vs multi-phase), the optimized prompt extracted 36 of 37 total compositions with only two FPs, yielding precision of 0.944 and recall of 0.919. This broader evaluation better reflects the actual behavior of the optimized prompt as specified in its instructions. 

Entity-level performance (\autoref{tab:compare_pr}) shows that nominal composition and processing condition maintained perfect precision (1.000), while lattice constant precision decreased slightly (0.912). Upon further analysis, we found that the reduction was caused by issues related to multi-phase handling. For instance, Zuo et al.~\cite{ZUO201460} investigated a series of Al$_\textrm{x}$CoFeNi HEA compositions. They found that when $\textrm{x}=0.5$ and $0.75$, the alloys exhibited a dual-phase microstructure consisting of both FCC and BCC phases (Table 1 in Zuo et al.~\cite{ZUO201460}). Additionally, in Table 2 (in Zuo et al.~\cite{ZUO201460}), the authors reported distinct lattice constants for the FCC and BCC phases corresponding to $\textrm{x}=0.5$ and $0.75$. However, the optimized prompt did not accurately associate each lattice parameter with its respective phase, which resulted in incorrect lattice constant extraction. This resulted in a slightly low precision for the lattice constants and phase in \autoref{tab:compare_pr}.

\begin{table}[ht!]
\centering
\caption{Comparison of entity-level extraction performance between the initial prompt and the optimized prompt using the Claude 3.5 Sonnet model.}
\label{tab:compare_pr}
\begin{tabular}{
  l
  S[table-format=1.3] 
  S[table-format=1.3] 
  S[table-format=1.3] 
  S[table-format=1.3] 
  S[table-format=1.3] 
  S[table-format=1.3] 
}
\toprule
\multirow{2}{*}{\textbf{Named Entity}} &
\multicolumn{3}{c}{\textbf{Initial prompt}} &
\multicolumn{3}{c}{\textbf{Optimized prompt}} \\
\cmidrule(lr){2-4}\cmidrule(lr){5-7}
& \textbf{Precision} & \textbf{Recall} & \textbf{F1 Score} & \textbf{Precision} & \textbf{Recall} & \textbf{F1 Score} \\
\midrule
Nominal composition   & 1.0 & 0.273 & 0.429 & 1.0 & 0.919 & 0.958\\
Lattice constant  & 1.0 & 0.273 & 0.429 & 0.912 & 0.838 & 0.873\\
Phase                 & 1.0 & 0.273 & 0.429 & 0.971 & 0.892 & 0.930\\
Alloy processing condition       & 1.0 & 0.273 & 0.429 & 1.0 & 0.919 & 0.958 \\
\bottomrule
\end{tabular}
\end{table}

\subsection*{Large-Scale Data Extraction}
\label{subsec:large_scale}
\begin{figure*}[!ht]
    \centering
    \includegraphics[width=\linewidth]{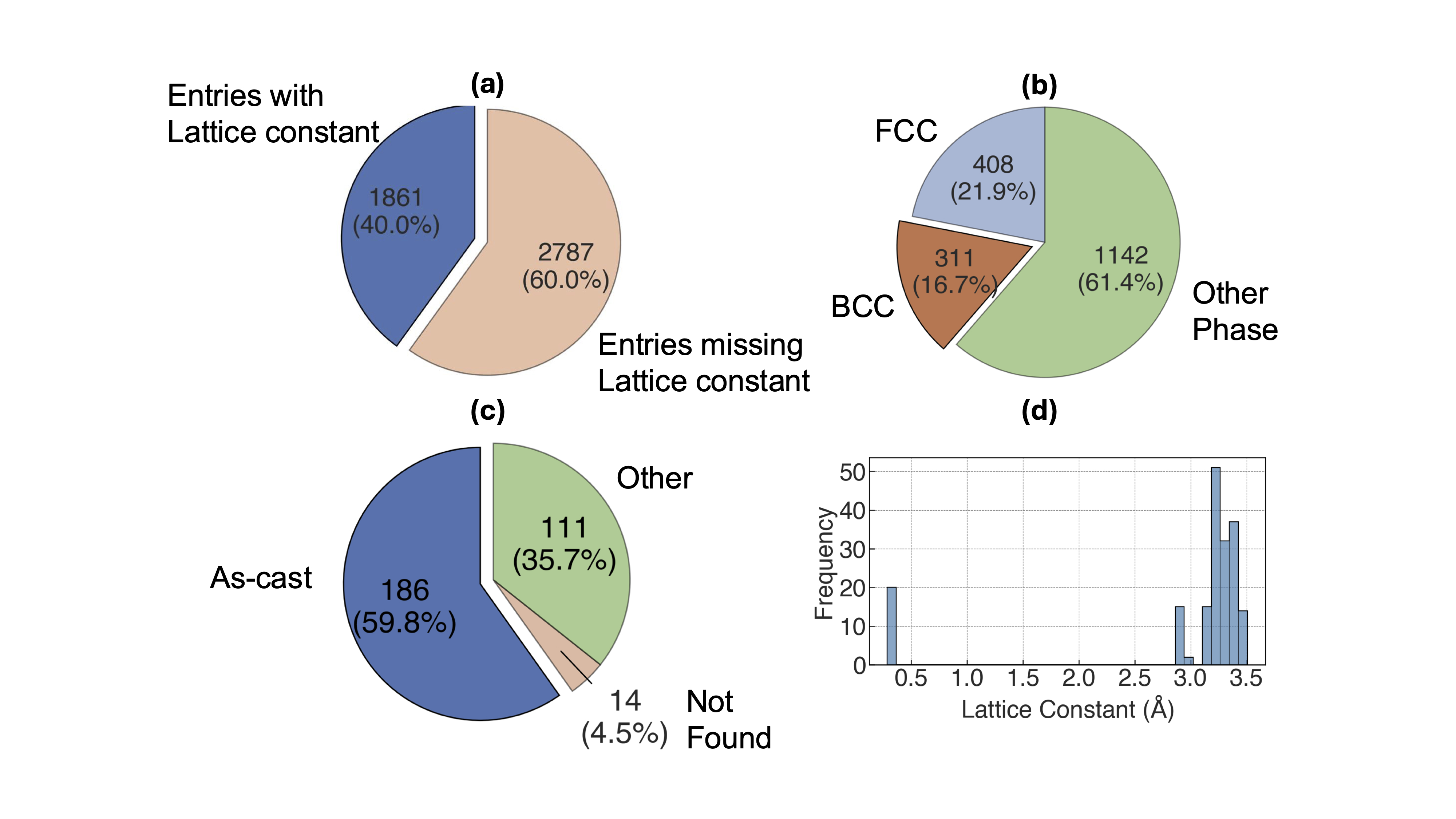}
    \caption{Overview of extracted HEA data from 2,267 publications. (a) A pie-chart showing the relative count of entries with and without lattice constant data: of 4,648 total extracted compositions, 1,861 (40.0\%) contained lattice constant while 2,787 (60.0\%) lacked this data. (b) A pie-chart showing the phase distribution among the 1,865 entries with lattice constants: 311 single-phase BCC (16.7\%), 408 single-phase FCC (21.9\%), and 1,142 other phases (61.4\%), including multi-phase mixtures, ordered structures, and amorphous structures. (c) A pie-chart showing the processing condition distribution for the 311 single-phase BCC alloys: 186 as-cast (59.8\%), 111 processed by other methods, such as selective laser melting and powder processing, or included post-processing heat treatments, such as annealing (35.7\%), and 14 with unreported processing information (4.5\%). (d) Histogram showing lattice constant distribution for the 186 as-cast single-phase BCC alloys. Most values fall within the range of 2.8–3.5~{\AA} for BCC structures. There were 21 entries near 0.3~{\AA} that represents a systematic unit conversion error where the LLM failed to convert nanometer values to {\AA} unit.}
    \label{fig:stat}
\end{figure*}

Out of the 2,267 publications that were processed through the Module 3 of the data extraction workflow (\autoref{fig:workflow}), 142 were rejected by the Claude 3.5 Sonnet model. These rejections occurred because these publications were either review articles that exceeded the context length limits or were classified as irrelevant to HEAs in the assessment conducted by the LLM. The large-scale data extraction module found a total of 4,648 alloy entries from 2,125 publications. Of these, 2,787 entries (60.0\%) contained no lattice constant data, while 1,861 entries (40.0\%) contained lattice constants, as visualized in \autoref{fig:stat}(a). Among the 1,861 entries with lattice constants data, 408 were face-centered cubic (FCC) alloys, 311 were BCC alloys, and 1,146 belonged to other phases (including multi-phase mixtures and other ordered structure types). This is shown in \autoref{fig:stat}(b). Out of the 311 single-phase BCC alloys, 186 were in the as-cast condition and 111 were processed or post-processed using other methods (e.g., powder processing, selective laser melting, annealing). Additionally, 14 entries lacked any reported processing information, as shown in \autoref{fig:stat}(c).

\autoref{fig:stat}(d) shows the lattice constants distribution for as-cast single-phase BCC alloys. The majority of values fall between 2.8 and 3.5~Angstroms~({\AA}), consistent with the expected BCC lattice parameters range. However, there were 21 entries with lattice constants around 0.3~{\AA}, indicating that the Claude 3.5 Sonnet model failed to convert lattice constants reported in nanometers to {\AA} during extraction. We identified a total of 327 entries with physically unreasonable lattice constants: 17 entries (from seven publications) with values $>$10~{\AA} and 310 entries (from 166 publications) with values $<$1~{\AA}. Manual inspection of all seven publications with anomalously large values plus ten randomly sampled papers with anomalously small values revealed that most outliers originated from lattice constants reported in nanometers or picometers in the source literature. Despite explicit instructions in the prompt to report values in {\AA}, the Claude 3.5 Sonnet model failed to perform the necessary unit conversions, representing a systematic formatting error. Along with the unit conversion errors, manual validation identified two additional categories representing systematic LLM extraction challenges: (1) contextual hallucination and (2) semantic misinterpretation. These challenges are summarized in \autoref{tab:llm_errors}, along with representative examples. 

\begin{table}[ht!]
\centering
\scriptsize
\caption{Summary of the three LLM scientific data extraction challenges that we encountered in this work.}
\begin{tabular}{@{}p{3.0cm} p{5.0cm} p{4.5cm}@{}}
\toprule
\textbf{Type} & \textbf{Explanation} & \textbf{Specific Example} \\ 
\midrule
\textbf{Contextual hallucination}
 & The model assigned a correct-looking number to the wrong entity due to mis-linking or mis-attribution. & The LLM mistakenly interpreted the lattice constant of the Ni$_{16}$Ti$_6$Si$_7$ precipitate phase as that of the host matrix, which has a different composition. (Yang et al.~\cite{YANG2023180}) \\ \\
\textbf{Semantic misinterpretation} & The model mixed up semantically related crystallographic parameters  because it lacked a strong understanding of their distinct physical meanings. & The LLM incorrectly identified Burgers vector as the lattice constant. (Moon et al.~\cite{MOON202155}) \\ \\
\textbf{Format or Unit conversion error} &  Skipped required unit conversion, despite explicit instructions in the prompt. & The LLM failed to convert units to Angstrom ({\AA}) in a few instances. (Piorunek et al.~\cite{PIORUNEK2020106792}, Jiang et al.~\cite{JIANG2022103136})  \\ 
\bottomrule
\end{tabular}
\label{tab:llm_errors}
\end{table}


To assess the performance of our data extraction, we randomly sampled 46 different publications, concentrating primarily on as-cast single-phase BCC alloy entries, which is our focus for the downstream task. We excluded outlier entries with anomalous lattice constants from this assessment. From the 46 publications sampled, we manually identified 44 as-cast single-phase BCC HEA compositions. Instead of treating phase and alloy processing conditions as separate entities, we combined them into a single composite criterion referred to as ``As-cast single-phase BCC.'' This simplification was essential because our current assessment focuses on alloys that exhibit a single-phase BCC structure under as-cast conditions. Therefore, we integrated phase identity and processing conditions into the composite criterion ``As-cast single-phase BCC'' to accurately represent this combined entity. This means if the LLM made a mistake on either phase or alloy processing condition, the entire entry was considered as FP or false negative (FN). For example, if the LLM extracted an as-cast dual-phase (BCC $+$ secondary phase) alloy but mislabeled it as single-phase BCC, the entire entry was counted as a FP. This entry contributed to the FP count when calculating precision and recall for the composite criterion ``As-cast single-phase BCC,'' and its associated nominal composition and lattice constant were also classified as FPs because they corresponded to a composition outside our scope or non-relevant. Conversely, if the LLM misclassified an as-cast single-phase BCC alloy as a dual-phase (BCC $+$ FCC) alloy, the entire entry was counted as a FN. Its nominal composition and lattice constant were classified as ``Not retrieved'' for our target category, regardless of whether those individual values were correctly extracted, because the alloy failed to meet our selection criteria.

Manual verification against the source publications showed high extraction accuracy in all fields, as measured by precision,  recall, and F1 score  (data given in \autoref{tab:llarge_scale_metrics}). Among the 47 extracted as-cast single-phase BCC entries, validation revealed five errors: three entries with incorrect alloy processing conditions, one computational result misclassified as experimental as-cast data, and one missed entry (FN). This yielded precision of 0.915 (43/47) and recall of 0.977 (43/44) for the composite criterion ``As-cast single-phase BCC.'' For the lattice constant field, we identified two additional errors, resulting in precision of 0.872 (41/47) and recall of 0.932 (41/44). The nominal composition field extraction performance matched the composite criterion exactly (precision = 0.915 and recall = 0.977). This indicates that the Claude 3.5 Sonnet model extracted compositions accurately whenever it correctly identified an entry as as-cast single-phase BCC. The precision and recall metrics use different denominators because precision is calculated as the ratio of true positives (TP) to the sum of TP and FP, whereas recall is calculated as the ratio of TP to the sum of TP and FN (additional details given in the Methods section).

\begin{table}[ht]
    \centering
     \caption{Precision and recall of the Claude 3.5 Sonnet LLM model were calculated through manual verification of a randomly sampled 46 publications. The named entities specified in the LLM prompt include the nominal composition, lattice constant, phase, and alloy processing conditions. Since we focused on HEA compositions that were processed as-cast and exhibit a single-phase BCC structure, we have combined phase and alloy processing conditions into one entity: As-cast single-phase BCC. We first filtered the extracted dataset to single-phase BCC alloys only, then calculated precision and recall metrics on the filtered dataset. It is important to note that the measured composition was not included in the assessment due to the complexity of its evaluation because not every publication reports measured composition.}
    \label{tab:llarge_scale_metrics}
    \begin{tabular}{l c c c}
        \toprule
        \textbf{Named Entity} & \textbf{Precision} & \textbf{Recall} &  \textbf{F1 Score}\\
        \midrule
        As-cast single-phase BCC & 0.915 & 0.977 & 0.945\\
        Nominal composition   & 0.915 & 0.977 & 0.945\\
        Lattice constant  & 0.872 & 0.932 & 0.901\\
        \bottomrule
    \end{tabular}
\end{table}

\subsection*{Outliers Discussion\label{sec:outlier}}
Out of the 17 extracted entries with lattice constants $>$10~{\AA}, two publications (Yang et al.~\cite{YANG2023180} and Pan et al.~\cite{PAN2022143692}) were not about HEAs, but were nevertheless returned by both the publisher API search and the Claude 3.5 Sonnet relevance classification step. Yang et al.~\cite{YANG2023180} investigated G-phase strengthened ferritic Fe-Cr model alloys with minor additions of Ni, Ti, and Si. The Claude 3.5 Sonnet model appears to have misclassified this as an HEA due to the presence of multiple constituent elements, demonstrating confusion between multi-component and HEA compositions. On closer examination, we found that the Claude 3.5 Sonnet had correctly recorded the alloy composition. However, it misunderstood the lattice constant of the Ni$_{16}$Ti$_6$Si$_7$ precipitate as that of the matrix. Interestingly, for this specific alloy, the LLM was able to perform unit conversion from nanometer to {\AA}, indicating that the error was not formatting-related but rather a semantic confusion about which phase the lattice constant described. This discussion is an example for contextual hallucination, where the LLM generates a plausible but incorrect extraction by combining information from different contexts. We confirmed this hallucination using the \texttt{EvaluatePdfOutputLoss} function. The loss function analysis (Figure S4 in SI) reveals that the LLM correctly identified the lattice constant as belonging to the G-phase, but incorrectly associated it with the bulk alloy composition. This indicates the LLM mistakenly attributed the lattice constant of the precipitate to the host alloy, failing to distinguish between matrix and precipitate phases.

A second contextual hallucination example occurred in Pan et al.~\cite{PAN2022143692}, who investigated a CoCrNiVC medium-entropy alloy. While the Claude 3.5 Sonnet correctly extracted the composition, phase, and processing condition, it erroneously reported the lattice constant as 10.9~{\AA}, which corresponds to that of the Cr$_{23}$C$_6$ precipitate and not the FCC matrix. Notably, the text loss analysis (Figure S5 in SI) revealed that the Claude 3.5 Sonnet recognized the lattice constant was associated with the carbide. However, it mistakenly attributed it to the alloy. This indicates that while the model was aware of the correct context, it failed to apply that knowledge accurately. A similar error occurred in Wu et al.~\cite{WU2017761}, where the Claude 3.5 Sonnet incorrectly extracted 11.303~{\AA} as the lattice constant of HEA matrix, when this value actually describes the Ti$_2$Ni intermetallic phase. The loss function analysis (Figure S6 in SI) confirmed this mis-attribution.

In one case, the Claude 3.5 Sonnet model misidentified a Burgers vector magnitude as a lattice constant. Moon et al.~\cite{MOON202155} investigated low-temperature deformation behavior of CoCrFeNiMo, but did not report lattice constants. Instead, the authors reported a Burgers vector magnitude of 0.255~nanometer (nm). The Claude 3.5 Sonnet model incorrectly extracted this value (converting to 2.55~{\AA}) as the HEA lattice constant. This error likely stems from the semantic similarity between Burgers vectors and lattice constants. The loss function analysis (Figure S7 in SI) confirmed that the model recognized this error: \textsf{``No, the value given (0.255) appears to be the Burgers vector magnitude, not the lattice constant.''}

Unit conversion failures represented a major source of outliers in the extracted dataset. Despite explicit prompt instructions to report lattice constants in {\AA}, the Claude 3.5 Sonnet model frequently extracted values in their original units (nanometers or picometers) without conversion during large-scale extraction. Two examples illustrate this systematic formatting error. In Piorunek et al.~\cite{PIORUNEK2020106792}, the source reported a lattice constant of 3.19~{\AA}, but the Claude 3.5 Sonnet model extracted 0.319~{\AA} (loss analysis is discussed in Figure S8 in SI). Similarly, for Jiang et al.~\cite{JIANG2022103136}, the Claude 3.5 Sonnet model extracted 0.323~{\AA} instead of 3.23~{\AA} (Figure S9 in SI). The loss function outputs revealed a concerning pattern: rather than acknowledging the unit conversion error, the Claude 3.5 Sonnet model insisted the extracted values (e.g., 0.319~{\AA}, 0.323~{\AA}) were correct in {\AA}. This represents a form of post-hoc rationalization where the model justifies an incorrect extraction rather than recognizing the error. Notably, the Jiang et al. case also illustrates a HEA relevance classification failure. The paper investigated Mg-Gd-Y-Zr alloys, not HEAs, yet passed both retrieval and extraction filters.

\subsection*{Testing Prompt Transferability}
\label{subsec:prompt_transfer}
To assess prompt transferability across models, we evaluated the initial unoptimized and optimized prompts on three recent commercial LLMs: ($i$) Claude 4.5 Sonnet, ($ii$) Gemini 2.5 Flash, and ($iii$) GPT-5. Claude 4.5 Sonnet and GPT-5 are models optimized for performance, whereas Gemini 2.5 Flash is a cheaper lightweight model optimized for efficiency and speed. Among these three new models, Gemini 2.5 Flash and GPT-5 enabled extended reasoning by default; on the contrary, the default setting of Claude 4.5 Sonnet is to disable extended reasoning. The default reasoning parameter in Gemini 2.5 Flash and GPT-5 were set as ``dynamic'' (\texttt{thinking\_budget=-1}), and ``medium'' (\texttt{reasoning\_effort="medium"}), respectively. We evaluated all three models on two distinct publication sets using their default temperature settings: ($i$) the seven-publication expert-labeled training set used for prompt optimization (Module 1, \autoref{fig:workflow}), containing 22 manually extracted single-phase HEA entries, and ($ii$) a 46-publication validation set randomly selected to assess large-scale extraction performance, containing 44 manually verified as-cast single-phase BCC HEA entries (discussed in the previous section). 
Performance metrics for extraction, specifically precision, recall, and F1 scores, are estimated for the initial prompt applied to the expert-labeled training set (Table S1 in the SI), as well as for the optimized prompt assessed on the 46-publication validation set (Table S2a and Table S2b in the SI). 
All three recent LLM models (with or without extended reasoning) significantly outperformed Claude 3.5 Sonnet on the expert-labeled training set, even when using the initial unoptimized prompt. Notably, Claude 4.5 Sonnet with the initial unoptimized prompt outperformed Claude 3.5 Sonnet with the fully optimized prompt on the expert-labeled training set. Similarly, GPT-5 with minimal reasoning and initial unoptimized prompt also achieved performance on the expert-labeled training set comparable to Claude 3.5 Sonnet model with the fully optimized prompt.

For the 46-publication validation set using optimized prompts, Claude 4.5 Sonnet without extended thinking achieved the best overall performance with zero FPs and four errors: one missed entry, one nominal composition error, and two lattice constant errors. It attained the highest F1 scores across all extracted fields, demonstrating that prompts optimized on Claude 3.5 Sonnet generalize effectively to newer models within the same LLM architecture family. Claude 4.5 Sonnet with extended reasoning achieved performance comparable to the default configuration, where the extended reasoning is disabled. While recall remained similar, precision decreased due to two FPs caused by over-interpretation of indirect information during extended reasoning. Additionally, the model missed two entries and produced one lattice constant error.

Gemini 2.5 Flash and GPT-5 showed similar performance, with F1 score differences of less than 0.04 across all fields. Gemini 2.5 Flash extracted 41 entries with two FPs, while GPT-5 produced 42 entries with three FPs and one nominal composition error. Both models achieved consistently high F1 scores despite differences in architecture, with only minor performance degradation relative to the Claude models. Although Gemini 2.5 Flash is a lightweight model compared to Claude Sonnet 4.5 and GPT-5, its performance is comparable to that of GPT-5, demonstrating successful prompt transferability across the LLM model families. The data given in Table S2a and Table S2b also includes the cost (in USD) for processing the 46-publication set with each LLM, along with the cost-effectiveness ratio (average F1 score per dollar). Higher ratios indicate better extraction performance per unit cost. Although Gemini 2.5 Flash had a relatively low average F1 score, it exhibited the best cost-effectiveness ratio. 

Among the FPs from Gemini 2.5 Flash and GPT-5, one notable case involved the TiVNbMoTa alloy from Zhang et al.~\cite{ZHANG2020155970}. In the original publication, the lattice constant of TiVNbMoTa was not explicitly reported by the authors but described through indirect comparison via average atomic volume. The Claude 3.5 Sonnet and Claude 4.5 Sonnet (without reasoning) models correctly recognized that the lattice constant was not directly provided and did not extract a value. In contrast, both Gemini 2.5 Flash (with dynamic reasoning) and GPT-5 (with medium reasoning) inferred values of 3.195~{\AA} and 3.194~{\AA}, respectively, for the TiVNbMoTa alloy. While these inferred values appear mathematically reasonable, suggesting extended reasoning capabilities, we classified them as FPs because they represent model inference rather than direct text extraction. 

To further confirm the role of reasoning, we re-ran the data extraction task on this specific publication~\cite{ZHANG2020155970} using Gemini 2.5 Flash and GPT-5 with various extended reasoning budgets. In the case of the Gemini 2.5 Flash model, when the dynamic reasoning capability was turned off (\textit{i.e.,} \texttt{thinking\_budget=0}), the model became consistent with the publication and reported the lattice constant as ``Not Found''. We repeated this test three times and the Gemini 2.5 Flash model consistently reported the lattice constant as ``Not Found''. When the thinking budget was set to default (dynamic reasoning), two independent runs returned the inferred result, while two additional runs reported the lattice constant as ``Not Found.'' This outcome is consistent with our expectation that extended reasoning can make responses less predictable but more robust because the model argues for multiple correct solutions. Upon reviewing the thought summaries of the output alongside the inferred results, we gained additional insights into the reasoning behind lattice constant inference. The extended reasoning for the Gemini 2.5 Flash model is given in Section S5 in the SI. The Gemini 2.5 Flash model begins by noting that the TiVNbTa and TiVNbMoTa alloys are discussed in the Results and Discussion section of the paper. Further along the reasoning process, the following sentence can be found: \textsf{``I see comparisons between the two, which provide a lattice constant for TiVNbMoTa.''}

In the minimal reasoning mode of the GPT-5 model, the performance was comparable to that of Gemini 2.5 Flash with dynamic reasoning; two out of four results indicated an inferred lattice constant. When the Claude 4.5 Sonnet model utilized extended reasoning with a token budget of 1024, its  performance became similar to both Gemini with dynamic reasoning and GPT-5 with medium reasoning. 
In one of the four re-runs, it showed an inferred value of 3.194~{\AA}, and two instances of thinking history explicitly detailed the calculations for this value. A representative example for the extended reasoning associated with the Claude 4.5 Sonnet model is given in Section S6 in the SI. Conversely, when the extended reasoning was disabled, the Claude 4.5 Sonnet model consistently returned ``Not found'' across three re-runs for this specific publication. This confirms that the presence of the inferred lattice parameter was a result of the extended reasoning process rather than a case of LLM hallucination.

The Table S2b gives the model performance of Claude 4.5 Sonnet, comparing optimized prompts with and without extended reasoning. Using a 1024-token budget for extended reasoning, the F1 scores for all entities showed a slight reduction, though they remained comparable. This observation suggests that, with an optimized prompt, the use of extended reasoning does not lead to further improvements in data extraction results.

Our comparison demonstrates that recent advancements in extended thinking and reasoning can significantly enhance performance on an unoptimized prompt. The reasoning process aims to analyze the intentions of the user and provide the expected result without the need for explicit prompt optimization. This approach proved effective for the relatively simple extraction task studied here, which involved only four primary entities (nominal composition, alloy processing condition, phase, and lattice constant). The validity of this result for more complex data extraction tasks (i.e., extraction of $>$10 named entities) requires additional work that is beyond the scope of this study. However, our case study also highlights a key risk: models with stronger reasoning capabilities may over-interpret indirect information. While such inferred values may appear plausible, they represent model interpretation rather than direct extraction from the source text. This observation reinforces the necessity of systematic prompt optimization frameworks, such as the \textsc{TextGrad}~\cite{yuksekgonul2025optimizing}, which can establish explicit constraints against over-interpretation. This ensures that advanced reasoning capabilities improve rather than compromise faithfulness to source material.

Overall, although frontier models achieve high accuracy even with the base prompt, this reinforces, rather than invalidates, the need for systematic prompt optimization. The Claude-optimized prompt improved the Gemini 2.5 Flash F1 score from 0.900 to 0.917 without re-optimization, demonstrating that optimization signals transfer across model families and model generation. Moreover, Gemini 2.5 Flash delivered similar accuracy at four to six times the cost efficiency of the other models. These results indicate that our prompt optimization distills model-agnostic features, such as unambiguous formatting rules, explicit handling for missing data, and clear scoping instructions, that generalize across model families and can be deployed flexibly, according to the available budget. Drawing on these results and observed failure modes, we identify four strategies to further improve extraction reliability. These strategies are discussed in Section S7 in the SI.

Furthermore, prompt optimization introduces a fixed, one-time cost that is easily amortized over large-scale deployments. In our workflow, optimization required only 21 article evaluations (seven articles across three epochs), which is approximately 1\% of the cost of the full 2,097-document extraction, yet yielded a 3$\times$ recall improvement in Claude 3.5 Sonnet. For small, exploratory tasks, direct use of the strongest available model remains cost-effective. However, for repeated or large-corpus extractions, it is almost certainly more economical to optimize on a capable mid-tier model and deploy on a lower-cost model, achieving comparable accuracy at substantially reduced expense.

\subsection*{Machine Learning for Predicting Lattice Constants}
The next step involves developing ML models to establish a quantitative relationship between lattice constants and HEA compositions of the as-cast, single-phase BCC alloys. 
The Figure S13a in the SI compares the ensemble of support vector regression (eSVR) model predictions against LLM-extracted experimentally measured lattice constant values, showing good agreement between the predicted and reported experimental data. Uncertainty for each composition was determined by calculating the standard deviation of the 100 individual SVR predictions that make up the ensemble, which serves as a measure of model confidence. The $R^2$ of training and test set were $0.989$ and $0.969$, respectively. 
Some scatter is evident, particularly in the lower and higher lattice constant regions ($\sim$2.8–3.0~{\AA} and $\sim$3.4-3.6~{\AA}, respectively). The larger uncertainty estimates in these ranges suggest either fewer training examples or greater compositional complexity relative to the mid-range values. Alloys in the lower lattice constant region were predominantly AlCoCrFeNiM$_\textrm{x}$-based compositions, where M is Si \cite{ZHU20107210}, Nb \cite{MA2012480}, or Mo \cite{ZHU20106975}. 
Similar to the eSVR models, the linear ensemble of LASSO models (eLASSO) also showed strong predictive performance. As shown in Figure S13b in the SI, the predicted lattice constants closely follow the experimentally measured values. The model uncertainty, estimated from the standard deviation of 1000 individual LASSO estimators, is small and therefore not visible on the plot. The eLASSO model achieved $R^2$ = 0.974 on both training and test sets, demonstrating good generalization performance.

\section*{Discussion}\label{sec:discussion}
We developed an expert-grounded, feedback-guided, automatic prompt optimization framework inspired by \textsc{TextGrad} and validated it using HEA lattice-constant extraction as a representative testbed. The workflow improved the recall of the Claude 3.5 Sonnet model from 0.27 to $>$~0.90 over only three epochs with seven hand-annotated documents, demonstrating that effective feedback-guided tuning can be achieved with minimal cost and human effort. The optimized prompt transferred across model families (Claude 4.5 Sonnet, GPT-5, and Gemini 2.5 Flash) with high fidelity, yielding modest additional gains on Gemini 2.5 Flash and enabling deployment on lower-cost architectures without re-optimization. These results establish feedback-guided prompt optimization as a scalable, transferable, and economically efficient methodology for future, more complex LLM-assisted scientific tasks.

When applied to a HEA lattice constant extraction problem, this approach enabled large-scale data retrieval from 2,267 publications. The optimized prompt achieved precision and recall exceeding 0.90 for as-cast single-phase BCC compositions, extracting lattice constant data for 1,861 HEA compositions. From this corpus, we curated a dataset of 166 as-cast single-phase BCC HEA compositions for ML validation. Models trained on this LLM-extracted data achieved test $R^2 > 0.95$ for lattice constant prediction, demonstrating that properly validated LLM-extracted data can support quantitative materials modeling with performance comparable to manually curated databases. 

While the optimized prompt transfers successfully across different LLM families, we observed that models with stronger reasoning capabilities occasionally introduced plausible but unreported values through over-inference. This behavior underscores that increasing reasoning effort does not guarantee epistemic reliability and reinforces the need for systematic validation protocols. 
Therefore, prompt optimization is not just about improving recall, but also about constraining model output to ensure faithfulness to the source material. 

Detailed error analysis further revealed three typical LLM failure modes: contextual hallucination, semantic misinterpretation, and unit-conversion errors. These highlight the continuing need for rigorous validation before quantitative use. Recommended safeguards include compositional consistency checks, physical plausibility filters, and random-sampling verification prior to downstream modeling. Looking ahead, extraction quality may be further improved by: ($i$) expanding the diversity of training documents, ($ii$) exploring limited forms of output-level optimization, ($iii$) favoring original publisher PDFs over XML reconstructions, and ($iv$) periodically reassessing performance with the most powerful current-generation frontier models. Further instructions for conducting any downstream task based on the data extracted from this work can be found in Section S9 in the SI.

Overall, this study demonstrates that prompt optimization can achieve considerable performance gains with minimal cost and effort; optimized prompts generalize across architectures yet require validation to prevent reasoning-induced over-inference; and validated LLM-extracted datasets can facilitate high-fidelity scientific data mining in the materials science domain.

\section*{Methods}\label{sec:methods}

\subsection*{Expert-Grounded Prompt Optimization Workflow}

The workflow for expert-grounded LLM-based data extraction is shown in \autoref{fig:workflow}. The approach comprises of three modules. The first module includes expert annotation of selected literature to define the target data structure and extraction schema that informs prompt engineering. For instance, to extract the lattice constant, we label each HEA with the following information: alloy name, nominal composition, measured composition, phase, alloy processing condition (such as as-cast, annealed, or powder processing), and the corresponding lattice constant. This information was compiled from a small selection of research papers. In this work, we considered seven publications (chosen randomly) and extracted 22 data points.

The second module performs automated, iterative prompt optimization via the \textsc{TextGrad} framework~\cite{yuksekgonul2025optimizing}. The pseudocode for the prompt optimization procedure is shown in Figure S14 in SI. Beginning with an initial natural language prompt describing the extraction task, we query a PDF-compatible multimodal LLM (Claude-3.5-Sonnet-20240620)~\cite{anthropic2024claude} to extract structured data from the same publications annotated by domain experts in Module 1. Extraction quality is assessed using a custom loss function, \texttt{EvaluatePdfExtractionLoss}, which computes the discrepancy between the LLM-generated output and expert-curated ground truth dataset from Module 1. The \texttt{EvaluatePdfExtractionLoss} function takes four inputs: the current prompt, scientific publication in PDF format, expert-constructed data, and the corresponding data generated by the LLM. The loss function uses a customized evaluation prompt (detailed in Figure S15 in SI) that instructs an LLM to assess whether the generated output aligns with both the extraction prompt requirements and the PDF content.  
After this analysis, it provides a clear statement indicating whether the generated data is in alignment with the prompt. The feedback generated by this function is then used by the backward engine to create modifications to the prompt, enabling an iterative refinement process. In this work, the optimization was performed with three epochs according to our budget for this module. Publications are batched in groups of three to remain within token limits. 

The third module performs large-scale data extraction across the HEA literature using the optimized prompt from Module 2. We applied this prompt to extract the following named entities: alloy name, nominal composition, measured composition, phase, alloy processing condition (e.g., as-cast, annealed, powder processing), and lattice constant. We used a total of 2,267 publications obtained from multiple sources, which include full-text XML files retrieved via Elsevier application programming interfaces (APIs) and PDF documents from other publishers. More specifically, we retrieved 2,093 full-text XML files and reconstructed to PDFs (text + table, no figures) from the digital object identifiers (DOIs) reported in Polak and Morgan's work~\cite{Polak2024}, and in addition, 174 downloaded full PDFs (text + table + figures) from various publishers. In Module 2, the Claude 3.5 Sonnet model used a default temperature value of 0.0 as specified in \textsc{TextGrad} for both forward and backward engine. In Module 3, we used the default temperature value of 1.0 with the Claude 3.5 Sonnet model. Similarly, for the Claude 4.5 Sonnet, Gemini 2.5 Flash, and GPT-5 models, we maintained the same default temperature of 1.0.

\subsection*{Data Extraction Performance Evaluation}
Our workflow requires two independent evaluation schemes to assess prompt performance: ($i$) validation against the expert-grounded reference dataset used during prompt optimization, and ($ii$) assessment of large-scale extraction quality after deployment on unseen publications. To quantitatively assess the performance, we calculated precision, recall, and F1 score using \autoref{eq:precision}, \autoref{eq:recall}, and \autoref{eq:f1} , respectively, based on the confusion matrix given in Table S3 in SI~\cite{wang2025, Manning_2008}.

\begin{align}
    \text{Precision} &= \frac{\text{Relevant Retrieved}}{\text{All Retrieved}}= \frac{\textrm{TP}}{\textrm{TP}+\textrm{FP}} \label{eq:precision}\\
    \text{Recall} &= \frac{\text{Relevant Retrieved}}{\text{All Relevant}}= \frac{\textrm{TP}}{\textrm{TP}+\textrm{FN}} \label{eq:recall} \\
    \text{F1}\ \text{Score}&= \frac{2\times \text{Precision}\times\text{Recall}}{\text{Precision} + \text{Recall}}
    \label{eq:f1}
\end{align}

Our HEA data extraction workflow functions on two hierarchical levels: ($i$) material identification and ($ii$) property extraction. For a retrieval to be classified as a True Positive (TP), both levels must be correct: the material must be an HEA discussed in the paper, \emph{and} its properties must be accurately extracted. If the identified material does not correspond to an HEA as referenced in the paper, all associated parameters are classified as False Positives (FP). This classification holds true even if the extracted property values are accurate.

In addition to precision and recall, we performed consistency checks to validate extraction quality prior to downstream tasks. These sanity checks verify agreement among three related fields: alloy name, nominal composition, and measured composition. These fields should exhibit consistency in constituent elements and approximate agreement in atomic ratios, as they represent different notations for the same material. We quantified agreement between composition fields using L1 distance ($D_{\mathrm{L1}}$ in \autoref{eq:l1}) and cosine similarity ($\mathrm{CosSim}$ in \autoref{eq:cosine}). In these equations, $\mathbf{v}_1$ and $\mathbf{v}_2$ represent composition vectors for the two fields being compared (e.g., nominal vs. measured), with each element $i$ corresponding to one of the $n$
unique constituent elements present in either composition.

\begin{align}
\label{eq:l1}
D_{\mathrm{L1}}(\mathbf{v}_1, \mathbf{v}_2)
&= \sum_{i=1}^{n} \left| v_{1,i} - v_{2,i} \right|, \\
\label{eq:cosine}
\mathrm{CosSim}(\mathbf{v}_1, \mathbf{v}_2)
&= 
\frac{\displaystyle \sum_{i=1}^{n} v_{1,i} v_{2,i}}
{\sqrt{\displaystyle \sum_{i=1}^{n} v_{1,i}^2} \,
 \sqrt{\displaystyle \sum_{i=1}^{n} v_{2,i}^2}}.
\end{align}

After calculating $D_{\mathrm{L1}}$ and $\mathrm{CosSim}$, we manually inspected entries with high compositional disagreement (large $D_{\mathrm{L1}}$, low $\mathrm{CosSim}$) among the alloy name, nominal composition, and measured composition fields, as these inconsistencies suggest material misidentification or incorrect data association. Separately, we excluded entries with lattice constants greater than 10~{\AA} and smaller than 1~{\AA}, which likely resulted from unit conversion errors during LLM data extraction.

We defined a second text loss function, \texttt{EvaluatePdfOutputLoss}, to assess the faithfulness of the extracted outputs. Inspired by Polak and Morgan's observation~\cite{Polak2024} that follow-up questions can improve extraction quality, we formulated several questions as validation instructions within the loss function (see Figure S16 in SI). This function takes both the LLM-generated output and the original PDF as inputs, then evaluates accuracy through fact-checking queries such as,  \textsf{``Are the lattice constants in the output truly the lattice constant of the material not others?''}. We primarily use this loss function to diagnose errors in outlier data, as discussed in Section S3 in the SI.

\subsection*{Downstream Task: Machine Learning To Predict Lattice Constants}
After data cleaning and quality control, 159 single-phase BCC HEA compositions were filtered to perform the downstream ML task. For each composition, we computed descriptors using the \textsc{Magpie} tool~\cite{ward2016general}, which generates composition-weighted averages of elemental properties (e.g., atomic volume, covalent radius). A descriptor selection pipeline based on recursive feature elimination method~\cite{Wang2023FeatureSIC, Sheikh2024ExploringCAB, Farrar2022MachineLAA} was implemented to identify the most important descriptors for training ML models (see Section S13 in SI). Feature selection identified six composition-weighted descriptors, which include: atomic volume (``meanAtomicVolume''), covalent radius (``meanColaventRadius''),  Mendeleev number (``meanMendeleev''), Electronegativity (``meanElectronegativity''), number of valence electron in the $d$-orbital (``meanNdValence''), and number of unfilled valence orbitals (``meanNUnfilled''). 

The dataset of 159 compositions was split into training and test sets using an 80:20 ratio (127 training, 32 test samples). We trained two ML models: ($i$) an ensemble of non-linear support vector regression (eSVR) models with radial basis function (RBF) kernels~\cite{smola2004tutorial, e1071, balachandran2020adaptive} and ($ii$) an ensemble of linear least absolute shrinkage and selection operator (eLASSO) models~\cite{tibshirani1996regression, elasso}. The hyperparameters (\textrm{gamma} and \textrm{cost}) for individual SVR estimator within the ensemble were independently optimized via grid search. The optimal ensemble size was selected from \{10, 20, 30, 40, 50, 75, 100\} estimators based on maximum test set $R^2$ performance.
The eLASSO model was trained using 1,000 bootstrap resamples that were generated from the training data. For each resample, the regularization parameter (shrinkage penalty) was optimized using grid search by selecting the value that minimized 10-fold cross-validation error. Both eSVR and eLASSO used identical train-test splits to enable direct performance comparison.

\bmhead{Supplementary information}

The [URL] will be inserted by the publisher.



\bmhead{Funding}
S.L. and P.V.B. acknowledge funding support from the University of Virginia Research Innovation Award. 

\bmhead{Acknowledgements}
S.L. and P.V.B. acknowledge funding support from the University of Virginia Research Innovation Award. Y.J. and P.V.B. thank the University of Virginia Research Interest Group initiative for fostering collaboration and exchange of ideas. S.L., T.R.B. and P.V.B thank Sunidhi Garg and Yifei Duan for insightful comments on the manuscript.

\section*{Author contribution}
The study was designed by S.L., Y.J., W.R., and P.V.B. The manuscript was prepared by S.L., T.R.B., Y.J., W.R., and P.V.B. The expert-constructed dataset was created by S.L. and T.R.B. under the supervision of P.V.B. Literature for data extraction was gathered by S.L., T.R.B., and P.V.B. The data extraction workflow was developed by S.L. with guidance from Y.J., W.R., and P.V.B. Manual verification of the extracted data was conducted by S.L., T.R.B., and P.V.B. All authors participated in discussing the results, writing, and providing feedback on the manuscript.

\section*{Competing interests}
The authors declare no competing interests.

\section*{Data Availability}
The expert-constructed dataset, extracted lattice constant dataset, the dataset sampled for evaluation in \autoref{tab:llarge_scale_metrics}, prompts, and text losses are archived on \texttt{Github} at \url{https://github.com/Shunl1996/expert-paper-extract} and provided with the paper. The DOIs for papers used in large-scale data extraction is provided as a dataset. Raw JSON outputs from LLMs for each paper, and thought summaries for selected papers are archived in the UVA Dataverse~\cite{raw_data} for reproducibility.

\section*{Code availability}
Codes for prompt optimization, large-scale data extraction, and composition verifications are available at \url{https://github.com/Shunl1996/expert-paper-extract} and also archived with the data described above.

\end{document}